\documentclass[12pt,english]{article}

\usepackage{hyperref}

\usepackage[T1]{fontenc}
\usepackage[latin9]{inputenc}

\usepackage{amsmath}
\usepackage{amssymb}

\makeatletter
\numberwithin{figure}{section}
\numberwithin{equation}{section}
\numberwithin{table}{section}

\newcommand{\be}{\begin{equation}}
\newcommand{\ba}{\begin{eqnarray}}
\newcommand{\ea}{\end{eqnarray}}
\newcommand{\ee}{\end{equation}}

\newcommand{\tha}{\theta}

\newcommand{\bea}{\begin{eqnarray}}
\newcommand{\eea}{\end{eqnarray}}
\newcommand{\bes}{\begin{equation*}}
\newcommand{\beas}{\begin{eqnarray*}}
\newcommand{\eeas}{\end{eqnarray*}}
\newcommand{\bas}{\begin{array*}}
\newcommand{\eas}{\end{array*}}
\newcommand{\ees}{\end{equation*}}
\newcommand{\nn}{\nonumber}
\newcommand{\p}{\partial}

\makeatother
\usepackage{comment}
\usepackage{babel}

\begin{document}
\begin{titlepage} \thispagestyle{empty}\begin{flushright}
YITP-22-50 \\
 
\par\end{flushright}

\bigskip{}

\noindent \begin{center}
{\textbf{\Large{}
On the Dynamics in the AdS/BCFT Correspondence}}\\
 \vspace{2cm}
 Yu-ki Suzuki and Seiji Terashima\vspace{1cm}
\par\end{center}

\noindent \begin{center}
\textit{  Center for Gravitational Physics and Quantum Information (CGPQI),}\\
\textit{ Yukawa Institute for Theoretical Physics, Kyoto University, 
}\\
\textit{ 
Kyoto 606-8502, Japan}\\
\par\end{center}

\begin{center}
\par\end{center}
\begin{abstract}
We consider a perturbation of the Einstein 
gravity with the Neumann boundary condition, which is regarded as
an end of the world brane (ETW brane) of the AdS/BCFT correspondence, 
in the 
$AdS_{d+1}$ spacetime with $d \geq 3$. 
We obtain the mode expansion of the perturbations explicitly for the tensionless ETW brane case.

We also show that 
the energy-momentum tensor in a $d$-dimensional BCFT should satisfy 
nontrivial constraints other than the ones for the boundary conformal symmetry
if the BCFT can couple to a $d$-dimensional gravity
with a specific boundary condition, which can be
the Neumann or the conformally Dirichlet boundary conditions.
We find these constraints are indeed satisfied for the free scalar BCFT.
For the BCFT in the AdS/BCFT, 
we find that
the BCFT can couple to 
the gravity with the Neumann boundary condition for the tensionless brane,  
but the BCFT can couple to the gravity with the conformally Dirichlet boundary condition
for the nonzero tension brane
by using holographic relations.


\end{abstract}
\end{titlepage}

\newpage{}

\tableofcontents{}

\newpage{}

\section{Introduction}

The Anti-de-Sitter/boundary conformal field theory (AdS/BCFT) correspondence \cite{Takayanagi:2011zk,Fujita:2011fp} is a generalization of AdS/CFT correspondence \cite{Maldacena:1997re}
in the sense that the space-time manifold admits a boundary. (See also \cite{Karch:2000gx}.)
There are
various works which confirms this duality in terms of the entanglement
entropy and the conformal anomaly both from the gravity sides and the BCFT sides, a small selection being \cite{Takayanagi:2011zk,Fujita:2011fp,Nozaki:2012qd,McAvity:1995zd,Collier:2021ngi,Kusuki:2021gpt,Numasawa:2022cni}. 
The AdS/BCFT plays an important role
in the recent studies of 
the replica wormhole and the island formula \cite{Almheiri:2019qdq,Penington:2019kki} 
in relation to the Page curve \cite{Page:1993df}.
 
In the AdS/BCFT correspondence we impose the Neumann boundary condition
on the metric on a boundary, which is called the end-of-the-world brane (ETW brane):
\begin{equation}
K_{ab}=(K-T)h_{ab},\label{eq:1}
\end{equation}
where the $K$ is an extrinsic curvature, $T$ is a constant, called the tension of the ETW brare, and $h_{ab}$
is an induced metric on the boundary.\footnote{
In this paper, the ETW brane simply means that 
we impose 
the boundary condition \eqref{eq:1} on a codimension one time-like hypersurface,
which ends on a fixed hypersurface in the asymptotic boundary of the AdS space,
and only consider one of the spacetime regions separated by the hypersurface.
} 
To see why we call it as Neumann boundary condition, we work
in the Gaussian normal coordinate where the boundary is on $r=0$ and
\begin{equation}
ds^{2}=dr^{2}+h_{ab}dx^{a}dx^{b}.
\label{eq:15}
\end{equation}
Then, we can see that 
\begin{equation}
K_{ab}=\frac{1}{2}\frac{\partial h_{ab}}{\partial r},\label{eq:16}
\end{equation}
which is the Neumann boundary condition on the metric. Here we note
that the indices $a,b$ run the tangential direction to an ETW brane. 
The gravity with boundary conditions have a long history, a small selection being \cite{Anderson:2006lqb,vanNieuwenhuizen:2005kg,Vassilevich:2003xt,Miao:2017gyt,Witten:2018lgb,Suzuki:2021pyw,Chu:2021mvq,Suzuki:2022xwv}. The meaning of fixing the extrinsic curvature as in the (\ref{eq:1}) can be found in \cite{Witten:2018lgb,Park:2015xoa}.

One of the important subjects of the AdS/BCFT correspondence is, of course, the dynamics, which have not been considered
even for the perturbation around the vacuum.
In the bulk gravity picture, this can be done by just studying the gravitational perturbations around the vacuum
with the boundary condition (\ref{eq:1}).

In this paper, we explicitly obtain the mode expansion of the gravitational perturbations around the AdS vacuum in the Poincare patch
with the boundary condition (\ref{eq:1}) for the tensionless case, $T=0$.
Using these results, we can obtain the bulk reconstruction formulas \cite{Bena} \cite{HKLL} assuming the BDHM relation \cite{Banks:1998dd} in principle, although we leave such work as a future problem. 
The bulk wave packets can be also considered in the AdS/BCFT correspondence following \cite{Te1,Te2}, which used an alternative bulk-boundary map given in \cite{Te3, Te4}.
This will be also interesting future work.

In this paper, we also show that if a $d$-dimensional BCFT can couple to a $d$-dimensional gravity with a boundary,
the energy-momentum tensor should satisfy nontrivial constraints other than the ones for the boundary conformal symmetry.\footnote{
Note that 
even if we consider the AdS/BCFT,
this $d$-dimensional gravity is different from the 
$d+1$-dimensional bulk gravity.
}
More precisely, we need to choose the boundary condition of the gravity.
We consider two known consistent boundary conditions,
namely, the Neumann and the conformally Dirichlet boundary conditions. We note that the Dirichlet boundary condition may be incosistent\cite{Witten:2018lgb}, therefore in this paper we do not consider the Dirichlet case.
We find that the necessary conditions for energy-momentum tensor in the BCFT which couples to the gravity with the Neumann boundary condition are
\begin{equation}
T_{x a}|_{x=0}=0,  \,\,\,
\partial_xT_{ab} |_{x=0}=0, \,\,\,
\partial_xT_{xx} |_{x=0}=0,
\label{nbci}
\end{equation}
which are satisfied for the free scalar BCFT with the Neumann and also the  Dirichlet boundary conditions.
The necessary conditions for energy-momentum tensor in the BCFT which couples to the gravity with the conformal Dirichlet boundary condition are
\begin{equation}
T_{x a}|_{x=0}=0,  \,\,\,
T_{ab} |_{x=0}= - \frac{1}{d-1} \eta_{ab} \, T_{xx}|_{x=0}, \,\,\,
\partial_xT_{xx} |_{x=0}=0,
\label{dbci}
\end{equation}
which are satisfied for the free scalar BCFT with the Dirichlet boundary condition.

For the BCFT in the AdS/BCFT, 
we find that
\eqref{nbci} are satisfied for the AdS/BCFT with tensionless ETW brane and
\eqref{dbci} are satisfied for the AdS/BCFT with non-zero tension ETW brane
by explicitly deriving the constraints for the energy-momentum tensor 
using holographic relations. 
Thus, 
the BCFT in the AdS/BCFT with the tensionless brane can couple to 
the gravity with the Neumann boundary condition, 
but the BCFT for the nonzero tension brane can couple to the gravity with the conformally Dirichlet boundary condition.
These results might be surprising because the ETW brane in the AdS/BCFT
imposes the Neumann boundary conditions for the bulk gravity theory,
even for the non-zero tension case.
We expect that the key to this problem is that we impose the conformally Dirichlet boundary conditions at the asymptotic AdS boundary. 
It will be interesting to understand these results clearly. We will expect that our results can also be examined by the holographic renormalization of the bulk geometry \cite{Guijosa:2022jdo}.

Note: This work has a overlap with \cite{Izumi:2022opi} partially. In that work they solve Einstein equation with the condition (\ref{eq:1}). Technically the assumptions they set are a little different from ours.

This paper is organized as follows:
In section 2 
we give a review of the mode expansion
of the metric in the AdS/CFT in the Poincare patch.
In section 3 we consider the mode expansion of the metric in AdS/BCFT
and obtain the explicit form of the mode expansion for the tensionless case.
In section 4 
we show that 
the energy-momentum tensor in a $d$-dimensional BCFT should satisfy 
nontrivial constraints 
if the BCFT can couple to a $d$-dimensional gravity
with a specific boundary condition.
For the BCFT in the AdS/BCFT, 
we find that
the BCFT can couple to 
the gravity with the Neumann boundary condition for the tensionless brane,  
but the BCFT can couple to the gravity with the conformally Dirichlet boundary condition
for the nonzero tension brane,
by using holographic relations. 
Section 5 is devoted to the conclusions.
In the Appendix, we consider the 
mode expansion
of the metric in the AdS/BCFT in the gauge invariant formalism,
although we only obtain partial results.

\section{ Free theory limit of gravity
in Poincare $AdS$ space} 
\label{s1}

In this section, in preparation for the study of the dynamical degrees of freedom of AdS/BCFT,
we will review the 
free theory limit of gravity
in Poincare $AdS_{d+1}$ space, which is supposed to be dual to 
the generalized free limit of the holographic CFT on the Minkowski space.\footnote{
Assuming the BDHM relation \cite{Banks:1998dd},
the reconstruction of the bulk local operators can be done in principle.}
This will be done by knowing
the mode expansion of the gravitational
perturbation $g_{\mu\nu}^{\left(1\right)}$ 
around the $AdS_{d+1}$
background $g_{\mu\nu}^{\left(0\right)}$ in the Poincare coordinate:
\begin{equation}
ds^{2}=g_{MN}^{\left(0\right)} d x^M dx^N
=\frac{dz^{2}+\eta_{\mu \nu} dx^\mu dx^\nu}{z^{2}},
\end{equation}
where $\mu,\nu=1,\ldots,d$. 
In this paper, we assume $d \geq 3$ because for $d=2$ there is no usual gravitons
in $AdS_3$ and there are only the boundary gravitons. 
The mode expansion for this was already done in \cite{He} \cite{Kabat} and 
we will follow their studies.

We will take the Fefferman-Graham coordinate, i.e. 
\begin{equation}
g_{z M}^{\left(1\right)}=0.
\end{equation}
Then, a part of the linearized Einstein equations with the normalizable condition for $z \rightarrow 0$ 
gives the following conditions:
\begin{equation}
g_{ \mu}^{\,\,\, \mu \left(1\right)}=0, \,\,\, \partial^\mu g_{\mu \nu}^{\left(1\right)}=0.
\end{equation}
The remaining linearized Einstein equations become
\begin{equation}
\left( \partial^\alpha \partial_\alpha+\partial_z^2 +\frac{5-d}{2} \partial_z- \frac{2(d-2)}{z^2}
\right)  g_{\mu \nu}^{\left(1\right)} =0.
\end{equation} 
These equations can be solved by the Bessel functions as 
\begin{equation}
g_{\mu \nu}^{\left(1\right)} =
\int d \omega d k_\mu d \zeta_{\mu \nu} \,  e^{i (\omega t+k_\mu x^\mu) } \, \zeta_{\mu \nu} \, z^{\frac{d}{2}-2}(a_{\omega, k,\zeta} J_{d/2}(\sqrt{ \omega^2 -k_\mu k^\mu
} \, \, z)+b_{\omega, k,\zeta}Y_{d/2}(\sqrt{ \omega^2 -k_\mu k^\mu
} \, \, z)),
\label{mode1}
\end{equation}
where 
the integrations are constrained with $\zeta_{\mu \nu}=\zeta_{\nu \mu},\zeta_{\mu}^{ \,\,\, \mu}=0, k^\mu \zeta_{\mu \nu}=0, \zeta_{\mu \nu} \zeta^{\mu \nu}=1$.
Because $J_{\nu}(Z) \sim Z^\nu$ and $Y_{\nu}(Z) \sim Z^{-\nu}$ near $z=0$,
the coefficients  $b_{\omega, k,\zeta} =0$ so that the field should be (delta-function) normalized. 
Furthermore, we need to take $a_{\omega, k,\zeta}=0$ if
$\omega^2 <\sum_{p=1}^{n+1} (k_p)^2$ because  $J_{\nu}(Z) \sim  e^{|Z|}$   in the limit
 $Z \rightarrow \infty$ for the region.
Finally, by the free field quantization, the coefficients $a_{\omega, k,\zeta}$ will be the creation operators or 
the annihilation operator for $\omega>0 $  or  $\omega < 0 $, respectively
with a suitable normalization constant factor.

\section{Gravitational perturbations in the AdS/BCFT }

In this section, we consider the mode expansion of the 
gravitational perturbations in the AdS/BCFT
with the boundary condition  (\ref{eq:1}) ,
\begin{equation}
K_{ab}=(K-T)h_{ab}.
\label{eq:1a}
\end{equation}
Note that this implies $K=T \frac{d}{d-1}$ and $K_{ab}=\frac{T}{d-1} h_{ab}$. 
The mode expansion has been done for the case without the boundary condition
in the previous section, and then we will search which combinations of 
the modes satisfy the boundary condition.

In order to specify the boundary condition explicitly,  
we first write the background $AdS_{d+1}$ metric as,
 \begin{equation}
     ds^2=d\eta^2+\frac{\cosh^2{\eta}}{\zeta^2} (d\zeta^2-dt^2+\sum_{i=1}^{n} dw_{i}^{2})
 \end{equation}
where $n=d-2$,
which becomes
the familiar form of the metric in the Poincare coordinate,  
 \begin{equation}
ds^{2}=g_{M N}^{\left(0\right)} d x^M dx^N
=\frac{dz^{2}-dt^{2}+dx^{2}+\sum_{i=1}^{n} dw_{i}^{2}}{z^{2}},
\end{equation}
by the following coordinate transformation:
  \begin{eqnarray}
  z&=&\frac{\zeta}{\cosh{\eta}},\nn\\
  x&=&\zeta\tanh{\eta}.
  \end{eqnarray}
For this background, the 
boundary condition (\ref{eq:1}) is satisfied by restricting the spacetime
to the region $-\infty<\eta< \eta^*$,
where $\eta^*$ is determined by the tension $T$ as $T=(d-1) \tanh \eta^*$.
In the usual coordinate, the boundary is at 
$x/z=\sinh \eta^*$.
In particular, for tensionless case $T=0$, $\eta^*=0$ which means that the boundary is at  $x=0$.

Note that, if we choose the Gaussian normal coordinate like (\ref{eq:15}), 
the boundary condition (\ref{eq:1}) becomes just a partial derivative \cite{Nozaki:2012qd},
 \begin{equation}
     \frac{\partial h_{ab}}{\p r}=\frac{2T}{d-1}h_{ab}=2 \tanh \eta^* h_{ab}.\label{eq:17}
 \end{equation}

Note also that in this paper  we require the asymptotic AdS boundary condition
where we can take the FG gauge and 
the metric which includes the perturbations near $z=0$ can be written by
 \begin{equation}
ds^{2}
=\frac{dz^{2}+h_{\mu \nu}(x^\mu,z)  dx^\mu dx^\nu }{z^{2}},
\end{equation}
where 
 \begin{equation}
h_{\mu \nu}(x^\mu,z)  dx^\mu dx^\nu = \eta_{\mu \nu} + T_{\mu \nu} (x^\rho) z^{d} +{\mathcal O}(z^{d+1}),
\end{equation}
and $T_{\mu \nu} (x^\rho) $ is the energy momentum tensor of the corresponding 
BCFT up to a constant.
We require this is valid on $\eta=\eta^*$.
In particular, this implies that $ T_{\mu \nu} (x^\rho)  |_{ x=0}$
is well-defined and does not depend on how we approach the corner point $z=x=0$.
Our requirement is different from the condition imposed in \cite{Izumi:2022opi} 
where $ T_{\mu \nu} (x^\rho)  |_{ x=0}$ is allowed to be divergent for some modes.

\subsection{
Tensionless case}

Let us consider the tensionless case.
The perturbation of the metric can be written as

\begin{eqnarray*}
g_{MN}=g_{M N}^{\left(0\right)}+g_{MN}^{\left(1\right)}.\\
\end{eqnarray*}
We will place the tensionless ETW brane at $x=0$ in this Poincare $AdS_{d+1}$
background $g_{M N}^{\left(0\right)}$. 
First we  assume that
the ETW brane is on $x=0$
even after including the perturbations of the metric.
We will reconsider the case where the ETW brane is not on $x=0$ later.

The unit normal vector for the surface is given by 
\begin{eqnarray*}
n_M=\delta_{x M} /\sqrt{(g^{-1})^{xx}}
\simeq\left(\frac{1}{z}+\frac{z g_{xx}^{(1)}}{2}\right) \delta_{x M},
\end{eqnarray*}
up to the first order of the perturbation.
The extrinsic curvature is defined by
\begin{eqnarray*}
K_{ab}=e_{a}^{M} e_{b}^{N}\nabla_{M}n_{N}
=\frac{\p n_b}{\p x^a} - \Gamma^M_{ab} n_M=- \Gamma^M_{ab} n_M \simeq -\frac{1}{z} \delta \Gamma^x_{ab},
\end{eqnarray*}
where $x^a=\{ t,z, w_i  \}$ 
and the projection tensor is 
\begin{eqnarray*}
e_{a}^{M}=\frac{\p x^M}{\p x^a}=\delta_{a}^{M}.
\end{eqnarray*}
Here, the deviation of the Christoffel symbol $\delta {\Gamma^{x}_{ab}}$ from the background metric in the linearized approximation is
\begin{eqnarray*}
\delta {\Gamma_{ab}^{x}} &=& \frac{1}{2} \delta g^{xM} \left(g_{Ma,b}+g_{Mb,a}-g_{ab,M} \right)
+\frac{1}{2} g^{x M} \left( \delta g_{Ma,b} + \delta g_{Mb,a} - \delta g_{ab,M} \right) \\
&=& z^3 \left( g^{(0)}_{bz} g^{(1)}_{xa} +g^{(0)}_{az} g^{(1)}_{xb} -g^{(0)}_{ab} g^{(1)}_{xz} \right)
+ \frac12 z^2 \left(  g^{(1)}_{xa,b} + g^{(1)}_{xb,a} - g^{(1)}_{xa,x} \right).
\end{eqnarray*}
Substituting these altogether, the boundary conditions in components become
\begin{eqnarray*}
&K_{tt}=-z\left(g_{xt,t}^{(1)}-\frac{1}{2}g_{tt,x}^{(1)}\right)-g^{(1)}_{xz}=0,\\
&K_{tz}=K_{zt}=-\frac{z}{2}\left(g_{xt,z}^{(1)}+g_{xz,t}^{(1)}-g_{zt,x}^{(1)}\right)-g^{(1)}_{xt}=0,\\
&K_{zz}=-z\left(g_{xz,z}^{(1)}-\frac{1}{2}g_{zz,x}^{(1)}\right)+g^{(1)}_{xz}=0,\\
&K_{tw_i}=K_{w_i t}=-\frac{z}{2}\left(g_{xw_i,t}^{(1)}+g_{xt,w_i}^{(1)}-g_{tw_i,x}^{(1)}\right)=0,\\
&K_{zw_i}=K_{w_iz}=-\frac{z}{2}\left(g_{xz,w_i}^{(1)}+g_{xw_i,z}^{(1)}-g_{zw_i,x}^{(1)}\right)-g^{(1)}_{xw_i}=0,\\
&K_{w_{i}w_{i}}=-\frac{z}{2}\left(2g_{xw_{i},w_{i}}^{(1)}-g_{w_{i}w_{i},x}^{(1)}\right)+g^{(1)}_{xz}=0,\\
&K_{w_{i}w_{j}}=-\frac{z}{2}\left( g_{xw_{i},w_{j}}^{(1)}+g_{xw_{j},w_{i}}^{(1)}-g_{w_{i}w_{j},x}^{(1)} \right) =0, \,\,\, (i \neq j),
\label{eq:-6}
\end{eqnarray*}
where they are evaluated on $x=0$ and
the trace of the extrinsic curvature $K$ was set to be zero since we focus on the tensionless
case. 

Next, we take the Fefferman-Graham gauge and 
substitute the mode expansions of the metric perturbations obtained in the previous section to
the above boundary conditions.
Note that 
the ETW brane is on $x=0$ by the assumption.
Then, from $K_{tz}=0$ we find $z g_{xt,z}^{(1)}+2 g_{xt}^{(1)}=0 $, which means $g_{xt}^{(1)} = z^{-2} G(t,w_i)$ on $x=0$.
This can not be possible for the normalizable mode, thus we find  $g^{(1)}_{xt} |_{x=0}= 0$.
With this and $K_{tt}=0$, we find $\frac{\p g^{(1)}_{tt}}{\p x} |_{x=0}= 0$.
Similarly, we find 
$g^{(1)}_{x w_i} |_{x=0}=\frac{\p g^{(1)}_{t w_i}}{\p x} |_{x=0}= \frac{\p g^{(1)}_{w_i w_j}}{\p x} |_{x=0}=0 $.

In summary, the boundary conditions in the Fefferman-Graham gauge $g^{(1)}_{z M}=0$ are 
$g^{(1)}_{x a} |_{x=0}=\frac{\p g^{(1)}_{a b}}{\p x} |_{x=0}=0 $.
Thus, we find that the modes which satisfy the boundary condition using the results without the boundary, \eqref{mode1}, 
are 
\begin{eqnarray*}
g_{\mu \nu}^{\left(1\right)}  &\sim &
\left(
 \cos (k_x x) \, \zeta^N_{\mu \nu}  
+ \sin (k_x x)  \, \zeta^D_{\mu \nu} 
\right) e^{i (\omega t+k_a x^a) } \, z^{\frac{d}{2}-2} J_{d/2}(\sqrt{ \omega^2 -k_\mu k^\mu
} \, \, z) 
,
\label{mode2}
\end{eqnarray*}
where 
$\zeta^N_{x \alpha} =\zeta^N_{\alpha x}=0 $ 
and
$\zeta^D_{\alpha \beta} =0$ 
for $\alpha \neq x$ and $\beta \neq x$.
Moreover,
$\zeta^N_{\mu \nu} $ and $\zeta^D_{\mu \nu} $ are symmetric traceless, which implies $\zeta^D_{x x}=0$, and they should satisfy
\begin{eqnarray*}
-k^x  \zeta^N_{x \nu}  
+ i k^a \zeta^D_{a \nu} =0, \,\,\,
k^x  \zeta^D_{x \nu}  
+ i k^a \zeta^N_{a \nu} =0,
\label{mode3}
\end{eqnarray*}
which come from $\p^\mu g_{\mu \nu}^{\left(1\right)} =0$.
The non-trivial parts of the equations of \eqref{mode3}
are $k^x  \zeta^D_{x b}  
+ i k^a \zeta^N_{a b} =0$ 
and $-k^x  \zeta^N_{x x}  
+ i k^a \zeta^D_{a x} =0$, which implies $k^a k^b \zeta^N_{a b} = (k^x )^2 \zeta^N_{x x }= -  (k^x )^2 \zeta^{N \,\, c}_{ \,\, \,\,c }$.

Note that we can regard the independent components of $\zeta^N$ and $ \zeta^D$ as $\zeta^{N}_{ab}$ constrained by 
$k^a k^b \zeta^N_{a b} = -  (k^x )^2 \zeta^{N \,\, c}_{ \,\, \,\,c }$.
Thus, the number of the degrees of freedom is $d(d-1)/2 -1=(d-2)(d+1)/2$ which is same as the one for the theory without the boundary, as expected because the gravitational waves far from the boundary will be same as ones for the theory without the boundary.
Note also that in the Gaussian normal coordinate, the boundary condition seems to be given by the Neumann boundary condition on the plane wave along $x$ direction only, however, we have chosen the different gauge and the mode expansions given here are different 
from that.

\subsubsection{No mode if the ETW brane is not on $x=0$}

We have assumed that the ETW brane is stretching on $x=0$ in the FG gauge.
This means that we searched the modes which satisfying this condition and
the modes we obtained might not  be complete.
Below we will consider the case that the ETW brane is not stretching on $x=0$ and 
will find that there are no additional modes, thus \eqref{mode2} are complete.

Let us assume that we pick 
the gauge choice above, and adding the perturbations. 
Then correspondingly the brane deviates a little from $x=0$ in general and we denote the hypersurface  of the brane as $F \equiv x-f(z,t,w_i)=0$
where $f(z,t,w_i)$ is the order of the perturbation. 
Let us linearize the boundary condition (\ref{eq:1}). 
The unit normal vector $n+ \delta n$ of the ETW brane is given by,
 \begin{equation}
n_M+\delta n_M= 
\frac{\p_M F}{\sqrt{ (g^{-1})^{MN} \p_M F \p_N F}},
\end{equation}
which implies
\begin{equation}
\delta n_M= 
-\frac{1}{z} \p_M f  + \delta_{x M} \frac1z \p_M f+ \cdots,
\end{equation}
where we keep the terms of the liner order of the perturbations.
Below, we will omit the higher order of the perturbations in the equations
for the notational simplicity.
The extrinsic curvature is defined by
\begin{eqnarray*}
K_{ab}=e_{a}^{M} e_{b}^{N} \nabla_{M} (n_{N}+ \delta n_N )
=\frac{\p (n_{b}+ \delta n_b )}{\p x^a} - \Gamma^M_{ab} (n_{M}+ \delta n_M )
+\p_a f  \delta_{b z}  /z^2  ,
\label{Kab}
\end{eqnarray*}
where $x^a=\{ t,z, w_i  \}$ 
and the projection tensor is 
\begin{eqnarray*}
e_{a}^{M}=\frac{\p x^M}{\p x^a}=\delta_{a}^{M} + \delta^{M}_{x} \p_a f.
\end{eqnarray*}
Here, we have approximated the terms comes from the deviation of the projection tensor, $  \p_a f   \nabla_{x} n_{b} +\p_b f   \nabla_{a} n_{x}= 
\p_a f  \delta_{b z}  /z^2  
$, in \eqref{Kab}
in the linear order in the perturbation.
Then, the deviation of it proportional to $\delta n$ is 
$\delta K_{ab}=\frac{\p  \delta n_b }{\p x^a} - \Gamma^M_{ab}  \delta n_M 
+\p_a f  \delta_{b z}  /z^2  $, and
we can calculate the extrinsic curvature as
\begin{eqnarray}
K_{zz}&=& g_{xz}^{(1)}-\frac{z}{2}\left(2 g_{xz,z}^{(1)}- g_{zz,x}^{(1)}\right)
-\frac{\p}{\p z} \left( \frac1z \frac{\partial f}{\partial z}\right)
,
\label{eq:21}
\end{eqnarray}
\begin{eqnarray}
K_{zt}=-z \frac{\p}{\p z} \left( \frac{1}{z^2}  \frac{\partial f}{\partial t}\right)-\frac{z}{2}\left( g_{xt,z}^{(1)}+ g_{xz,t}^{(1)}- g_{zt,x}^{(1)}\right)-g_{xt}^{(1)},\label{eq:20}
\end{eqnarray}
\begin{eqnarray}
K_{zw_i}=-\frac{1}{z}\left(\frac{\partial^2f}{\partial w_i\partial z}\right)
-\frac{z}{2}\left( g_{xw_i,z}^{(1)}+ g_{xz,w_i}^{(1)}- g_{zw_i,x}^{(1)}\right)-g_{xw_i}^{(1)},
\end{eqnarray}
\begin{eqnarray}
K_{tt}&=& g_{xz}^{(1)}-\frac{z}{2}\left(2 g_{xt,t}^{(1)}- g_{tt,x}^{(1)}\right)
-\frac{1}{z}\left(\frac{\partial^2f}{\partial t^2}\right)-\frac{1}{z^2}\left(\frac{\partial f}{\partial z}\right),\label{eq:22}
\end{eqnarray}
\begin{eqnarray}
K_{tw_i}=-\frac1z \frac{\partial^2f}{\partial t\partial w_i}-\frac{z}{2}\left( g_{xt,w_i}^{(1)}+ g_{xw_i,t}^{(1)}- g_{w_it,x}^{(1)}\right),
\end{eqnarray}
\begin{eqnarray}
K_{w_iw_i}&=& -g_{xz}^{(1)}-\frac{z}{2}\left(2 g_{xw_i,w_i}^{(1)}- g_{w_iw_i,x}^{(1)}\right)
-\frac{1}{z}\left(\frac{\partial^2f}{\partial w_i^2}\right)+\frac{1}{z^2}\left(\frac{\partial f}{\partial z}\right),
\end{eqnarray}
\begin{eqnarray}
K_{w_iw_j}=-\frac1z \frac{\partial^2f}{\partial w_i\partial w_j}-\frac{z}{2}\left( g_{xw_i,w_j}^{(1)}+ g_{xw_j,w_i}^{(1)}- g_{w_iw_j,x}^{(1)}\right),
\label{eq:25}
\end{eqnarray}
where we have used the fact that non-zero $\Gamma_{MN}^P$ are
\begin{eqnarray*}
\frac{1}{z}=\Gamma_{ii}^z \,\, =-\Gamma_{tt}^z \,\, =-\Gamma_{zz}^z \,\, 
=-\Gamma_{iz}^i \,\, =-\Gamma_{zi}^i \,\, =\Gamma_{tz}^t \,\, =\Gamma_{zt}^t \,\, 
\end{eqnarray*}
for the Poincare AdS space in the coordinate system.

The boundary conditions are $K_{ab}|_{x=f}=0$.
In the Fefferman-Graham gauge $g^{(1)}_{z M}=0$, 
$K_{zz}|_{x=f}=0$ implies $f =c_1 z^2+ c_0$.
Here, $c_0$ should be zero  
because of 
the asymptotic boundary condition $f \rightarrow 0$ for $z \rightarrow 0$.
Furthermore, 
$K_{tt}|_{x=f}$ contains $-\frac{1}{z}\left(\frac{\partial^2f}{\partial t^2}\right)-\frac{1}{z^2}\left(\frac{\partial f}{\partial z}\right)= -2 c_1/z+{\cal O}(z)$,
which should be canceled by other terms.
Thus, we find $f=0$ and 
then we conclude that
there are no additional modes with $f \neq 0$.

\subsection{Non-zero tension case}
 
In this subsection we consider the non-zero tension case although 
we can not obtain a closed form of the mode expansion.

Let us consider the perturbations in the FG gauge, $g_{z M}^{(1)}=0$, and we denote the position of the brane as 
$F \equiv x-k z-f(z,t,w_i)=0$, where $k =\sinh \eta^*$. 
We note that $f(z,t,w_i)$ is the order of the perturbation ${\cal O}(g^{(1)})$.
The unit normal vector $n$ of the ETW brane is given by,
 \begin{equation}
n_M= 
\frac{\p_M F}{\sqrt{ (g^{-1})^{MN} \p_M F \p_N F}},
\end{equation}
where $\p_M F= \delta_{x M}-k \delta_{z M}-\p_M f$ and
\begin{eqnarray*}
(g^{-1})^{M N} \p_M F \p_N F =z^2(1+k^2){\eta^*}+2z^2k\p_z f-z^4g_{xx}^{(1)}
\end{eqnarray*}
Then, we find
\begin{equation}
n_M= 
\frac{1}{z\sqrt{1+k^2}} \left(1-\frac{k}{1+k^2}\p_z f+\frac{z^2}{2(1+k^2)}g_{xx}^{(1)}\right)\left(\delta_{x M}-k \delta_{z M}-\p_M f\right)
\end{equation}
In particular, we obtain
\begin{eqnarray}
n_z&=&\frac{1}{z\sqrt{1+k^2}}(-k
-\frac{1}{1+k^2}\p_z f-\frac{kz^2}{2(1+k^2)}g_{xx}^{(1)}),\nn\\
n_x&=& \frac{1}{z\sqrt{1+k^2}}(1-\frac{k}{(1+k^2)}\p_z f+\frac{z^2}{2(1+k^2)}g_{xx}^{(1)}),\nn\\
n_t&=&-\frac{1}{z\sqrt{1+k^2}}\p_t f,\nn\\
n_{w_i}&=&-\frac{1}{z\sqrt{1+k^2}}\p_{w_i} f.
\end{eqnarray}

The extrinsic curvature is defined by 
\begin{eqnarray*}
K_{ab}&=&e_{a}^{M} e_{b}^{N} \nabla_{M} n_{N}\nn\\
&=&\frac{\p n_{b}}{\p x^a} - \Gamma^M_{ab} n_{M}
+ (k \delta_{a}^{z}+\p_a f)  \nabla_{x} n_{b} + (k  \delta_{b}^{z} +\p_b f ) \nabla_{a} n_{x}+(k \delta_{a}^{z}+\p_a f) (k  \delta_{b}^{z} +\p_b f )\nabla_{x} n_{x}   ,
\end{eqnarray*}
where $x^M=\{ x=k z +f( t,z, w_i  ), x^a=\{ t,z, w_i  \}\}$ and the projection tensor is 
\begin{eqnarray*}
e_{a}^{M}=\frac{\p x^M}{\p x^a}=\delta_{a}^{M}+ k \delta_{x}^{M} \delta_{a}^{z}   + \delta^{M}_{x} \p_a f.
\end{eqnarray*}
The boundary condition for the non-zero tension case is 
\begin{eqnarray}
K_{ab} &=&\tanh \eta^*  h_{ab},
\label{bcg}
\end{eqnarray}
and we will firstly compute the $zz$ component.
This is evaluated as
\begin{eqnarray*}
K_{zz}=\frac{1}{\sqrt{1+k^2}}\left(\frac{k}{z^2}+\frac{k^3}{z^2}-\frac{1}{z}\p^2_z f+\frac{1}{z^2}\frac{2k^4+3k^2+1}{k^2+1}\p_z f-\frac{3k}{2}g_{xx}^{(1)}-g_{xx,z}^{(1)}(zk+\frac{zk^3}{2})-g_{xx,x}^{(1)}\frac{zk^2}{2}\right).
\end{eqnarray*}
Similarly we can expand the right hand side of (\ref{bcg})
and we obtain
\begin{equation}
    \tanh \eta^* h_{zz}=\frac{1}{\sqrt{1+k^2}}\left(\frac{k}{z^2}+\frac{k^3}{z^2}+k^3g_{xx}^{(1)}+\frac{2k^2}{z^2}\p_z f\right),
\end{equation}
which implies that $f$ is ${\cal O}(z^{d+1})$  and completely determined
by 
\begin{eqnarray}
\frac{1}{z}\p^2_z f - \frac{1}{z^2}\p_z f-\frac{(2k^2+3)k}{2}g_{xx}^{(1)}-g_{xx,z}^{(1)}(zk+\frac{zk^3}{2})=0,
\end{eqnarray}
where we omit $zg_{xx,x}^{(1)}$ because it is the order of $z^{d-1}$.
In particular, for $g^{(1)}_{xx} = T_{xx} z^{d-2} + {\cal O}(z^{d-1})$,
we find 
\begin{eqnarray}
f= -\frac{dk^3+(2d-1)k}{2(d^2-1)}T_{xx}z^{d+1}  + {\cal O}(z^{d+2}).
\label{fz}
\end{eqnarray}
We note that there could be ${\cal O}(z^{2})$ or ${\cal O}(z^{0})$ terms in $f$ 
other than the those determined by $g_{xx}^{(1)}$.
However, such terns are forbidden by 
the other components of the boundary conditions as we will see below.
Note that  ${\cal O}(z^{0})$ term is not allowed because $f$ should become zero in the limit
$z \rightarrow 0$ because the ETW brane is should be terminated at $x=z=0$.

For the boundary conditions for other components, we obtain
\begin{eqnarray}
zt&:&\frac{1}{\sqrt{1+k^2}}\left(\frac{k^2}{z^2}\p_t f-\frac{1}{z}\p_z\p_t f-\frac{z(1+k^2)}{2}g_{xt,z}^{(1)}-(1+k^2)g_{xt}^{(1)}-\frac{zk}{2}g_{xx,t}^{(1)}\right)=0,\nn\\
zw_i&:&\frac{1}{\sqrt{1+k^2}}\left(\frac{k^2}{z^2}\p_{w_i} f-\frac{1}{z}\p_z\p_{w_i} f-\frac{z(1+k^2)}{2}g_{xw_i,z}^{(1)}-(1+k^2)g_{xw_i}^{(1)}-\frac{zk}{2}g_{xx,w_i}^{(1)}\right)=0,\nn\\
tt&:&\frac{1}{\sqrt{1+k^2}}\left(-\frac{1}{z^2(1+k^2)}\p_z f-\frac{1}{z}\p_t^2 f-\frac{k}{2(1+k^2)}g_{xx}^{(1)}-\frac{zk}{2}g_{tt,z}^{(1)}+\frac{z}{2}g_{tt,x}^{(1)}-kg_{tt}^{(1)}-zg_{xt,t}^{(1)}\right)=0,\nn\\
tw_i&:&\frac{1}{\sqrt{1+k^2}}\left(-\frac{1}{z}\p_{w_i}\p_t f-\frac{zk}{2}g_{w_it,z}^{(1)}-kg_{tw_i}^{(1)}-\frac{z}{2}(g_{xt,w_i}^{(1)}+g_{xw_i,t}^{(1)}-g_{tw_i,x}^{(1)})\right)=0,\nn\\
w_iw_i&:&\!\frac{1}{\sqrt{1+k^2}}\!
\left(\frac{1}{z^2(1+k^2)}\p_z f-\!\frac{1}{z}\p_{w_i}^2 f+\!\frac{k}{2(1+k^2)}g_{xx}^{(1)}
\right. 
\nn\\
&& \!\ -
\left.
\frac{zk}{2}g_{w_iw_i,z}^{(1)}+\!\frac{z}{2}g_{w_iw_i,x}^{(1)}-\!kg_{w_iw_i}^{(1)}-\!zg_{xw_i,w_i}^{(1)} 
\right) =0,
\nn\\
w_iw_j&:&\frac{1}{\sqrt{1+k^2}}\left(-\frac{1}{z}\p_{w_i}\p_{w_j} f-\frac{zk}{2}g_{w_iw_j,z}^{(1)}-kg_{w_iw_j}^{(1)}-\frac{z}{2}(g_{xw_j,w_i}^{(1)}+g_{xw_i,w_j}^{(1)}-g_{w_jw_i,x}^{(1)})\right)=0.
\label{bcf1}
\end{eqnarray}
If $f= \alpha z^2+ \cdots$,
these are dominant in $z \rightarrow 0$
in the above $tt$ component and we find
$-\frac{1}{(1+k^2)} 2 \alpha -\p_t^2 \alpha z^2=0$, which implies $\alpha=0$

We would like to find the mode which satisfies the boundary conditions.
Note that each mode of the $g_{M N}^{(1)}$ is proportional to the Bessel function 
$J_{d/2}(\sqrt{\omega^2 - k^\mu k_\mu } z)$ in the Fefferman-Graham gauge, however, $f$ is completely different from the Bessel function. 
Thus, in order to solve the boundary conditions generically we need to 
consider linear combinations of infinitely many modes with different $k_\mu$, but same $\omega$.
In principle, there will be only the technical problem to do this, however,
in this paper we leaves this problem open for completion in the future.

Finally, in order to obtain the constraints for the energy-momentum tensor,
we can extract the coefficient of $z^{d-2}$ in the above equations \eqref{bcf1}. Then we obtain
\begin{equation}
T_{xt}=T_{xw_i}=T_{tw_i}=T_{w_iw_j}=0,
\end{equation}
where we define
\begin{equation}
g^{(1)}_{ij} = T_{ij} z^{d-2} + {\cal O}(z^{d-1}).
\end{equation}
For the $tt$ and $w_iw_i$ components we can substitute the form of (\ref{fz}) we find that 
\begin{eqnarray}
T_{tt}&=&\frac{1}{d-1}T_{xx},\\
T_{w_iw_i}&=&-\frac{1}{d-1}T_{xx}.
\end{eqnarray}
Therefore we can combine these conditions as 
\begin{equation}
    T_{ij}=-\frac{1}{d-1}\eta_{ij}T_{xx},
\end{equation}
where the $\eta_{ij}$ is the Minkowski metric. 
We will see what these conditions mean in the next section.

\label{s3}

\section{Constraints of the energy-momentum tensor in BCFT }

In this section, we consider general BCFT in flat space and derive constraints of the energy-momentum tensor of it by considering the coupling to the gravity.
We will also find that the energy-momentum tensor in AdS/BCFT satisfies the constraints.

First, by the definition of the $d$-dimensional BCFT, the boundary conformal symmetry should be kept, namely, 
the presence of the boundary breaks the conformal symmetry $SO(2,d)$ to $SO(2,d-1)$, which means 
the condition on the following energy-momentum tensor:
\begin{equation}
T_{x a}|_{x=0}=0,  
\label{cbc}
\end{equation}
where we put the boundary on the $x=0$ surface and 
$a,b,\cdots$ run the transverse directions to the boundary surface $x=0$.
Note that \eqref{cbc} implies
\begin{equation}
\p_x T_{x x}|_{x=0}=0,
\label{cbc2}
\end{equation}
by the current conservation $\p^x T_{xx}+\p^a T_{xa}=0$.
This boundary condition requires that the energy flux
does not spread out from the boundary of the CFT region. This is just
a generalization of the two-dimensional case $(T-\bar{T})|_{x=0}=0$ \cite{Cardy:1984bb}.
 


Then, we will consider a $d$-dimensional BCFT which can couple to a $d$-dimensional  gravity theory with the boundary
and the necessary conditions for the properties of such a BCFT.
Here, we need to specify the boundary condition of the gravity theory\footnote{
Similar considerations were given in \cite{Prochazka:2018bpb}.}
and we will choose the Neumann boundary condition for simplicity
and later we will consider the conformal Dirichlet boundary condition.\footnote{
The Dirichlet  boundary condition may be inconsistent \cite{Witten:2018lgb}.
There might be other types of consistent boundary conditions, which are not considered in this paper.
}
In order to fix the location of the boundary in the BCFT,
we choose the Gaussian normal coordinate so that the boundary is fixed at $x=0$. Then, the Neumann boundary condition means $\partial_{x}g_{ab}|_{x=0}=0$  and,
using Fourier transformation, the metric can be represented as 
\begin{equation}
    \delta g_{ab}=g_{ab} -\eta_{ab} \sim\int dk f_{ab}(k)\cos{(kx)}.
\end{equation}
Thus, in order to couple the gravity consistently,\footnote{
Here, we assume that the nonzero energy momentum tensor should be a source for the gravity. For a system which has a subsystem which 
is completely decoupled from the other subsystem and the gravity,
we can consider only the subsystem which couples to the gravity.
We thank M. Shigemori for the useful discussions on this point.
} 
the energy-momentum tensor in the BCFT should also satisfy 
the Neumann boundary condition,
\begin{equation}
    \partial_xT_{ab} |_{x=0}=0,
\label{cons1}
\end{equation}
because of
the coupling of the energy-momentum tensor to the perturbations of the metric $\delta g$:
 \begin{equation}
\int T_{\mu\nu} \delta g^{\mu\nu}.
\label{eq:60}
 \end{equation}
(The gauge dependence is of course absent by the current conservation.)
Using the traceless condition,
we find that the necessary conditions for energy-momentum tensor in the BCFT which couples to the gravity with the Neumann boundary condition are
\begin{equation}
T_{x a}|_{x=0}=0,  \,\,\,
\partial_xT_{ab} |_{x=0}=0, \,\,\,
\partial_xT_{xx} |_{x=0}=0.
\label{nbc}
\end{equation}
Note that for two dimensional BCFT, 
$T_{x a}|_{x=0}=0$ implies others by using the current conservation
$\p_x T_{xx} -\p_t T_{tx}=0$ and the traceless condition.

Although a BCFT does not need to satisfy the boundary conditions \eqref{nbc}, we will see that the BCFT which is realized as a free scalar field theory 
with a boundary indeed satisfy them.
Let us consider a free scalar field on a half of $\mathbb{R}^{1,d-1}$. 
Here we set the boundary at $x=0$ and we impose 
the Neumann or Dirichlet boundary conditions at this boundary. 
The Lagrangian of the conformally coupled scalar filed 
is
\begin{equation}
    L=-\frac{1}{2}\sqrt{-g}\left(g^{\mu\nu}\partial_{\mu}\phi\partial_{\nu}\phi-\xi R \phi^2\right),
\end{equation}
where $\xi=\frac{1}{4}\frac{d-2}{d-1}$. The energy-momentum tensor on general background is given by
\begin{eqnarray}
T_{\mu\nu}&=&\frac{2}{\sqrt{-g}}\frac{\delta S}{\delta g^{\mu\nu}}\nn\\
&=&(1-2\xi)\partial_{\mu}\phi\partial_{\nu}\phi+(2\xi-\frac{1}{2})g_{\mu\nu}g^{\rho\sigma}\partial_{\rho}\phi\partial_{\sigma}\phi-2\xi\phi\partial_{\mu}\partial_{\nu}\phi+\frac{2}{d}\xi g_{\mu\nu}\phi\Box\phi\nn\\
&-&\xi\left(R_{\mu\nu}-\frac{1}{2}Rg_{\mu\nu}+\frac{2(d-1)}{d}\xi Rg_{\mu\nu}\right)\phi^2.
\end{eqnarray}
For the flat space , we obtain
\begin{equation}
   T_{\mu\nu}=\partial_{\mu}\phi\partial_{\nu}\phi-\frac{1}{2} \eta_{\mu\nu}\partial^{\alpha}\phi\partial_{\alpha}\phi+\frac{d-2}{4(d-1)}\left((\eta_{\mu\nu}\partial^{\alpha}\partial_{\alpha}-\partial_{\mu}\partial_{\nu})\phi^2+(\frac{2}{d}-2) \eta_{\mu\nu}\phi\Box\phi\right).
\end{equation}
The last term proportional to  $\Box\phi$ vanishes by the equations of motion and this is equivalent to the usual improved energy-momentum tensor.
Then, for the Neumann boundary condition $\p_x \phi|_{x=0}=0$,
which implies $\p_a \p_x \phi|_{x=0}=0$, we can easily see that 
\eqref{nbc} are satisfied using $\partial^{\alpha}\partial_{\alpha} \phi=0$.
For the Dirichlet boundary condition $\phi|_{x=0}=const.$,
which implies $\p_a  \phi|_{x=0}=0$, we can easily see\footnote{
Note that $\partial^{x}\partial_{x} \phi |_{x=0} =-\partial^{a}\partial_{a} \phi |_{x=0} =0$.
}
that \eqref{nbc} is satisfied if we require $\phi|_{x=0}=0$ which is obviously needed for 
the conformal symmetry.

Next, we will 
consider a $d$-dimensional BCFT which can couple to a $d$-dimensional  gravity theory with the conformally Dirichlet boundary \cite{Anderson:2006lqb}.
For this, the metric fluctuations should be $\delta g_{a b} |_{x=0} \sim \eta_{a b}$,
which implies $T_{ab} |_{x=0} \sim \eta_{a b}$.
Thus, 
the necessary conditions for energy-momentum tensor in the BCFT which couples to the gravity with the conformally Dirichlet boundary condition are
\begin{equation}
T_{x a}|_{x=0}=0,  \,\,\,
T_{ab} |_{x=0}= - \frac{1}{d-1} \eta_{ab} \, T_{xx}|_{x=0}, \,\,\,
\partial_xT_{xx} |_{x=0}=0.
\label{dbc}
\end{equation}
where we have used the traceless condition.
For the free scalar field theory, we can see that 
these conditions are not
satisfied for the Neumann boundary condition,
but are satisfied for the Dirichlet boundary condition.

\subsection{Constraints of the energy-momentum tensor in AdS/BCFT }

We will study the constraints of the energy-momentum tensor in AdS/BCFT,
where the Neumann boundary condition is imposed on the bulk $(d+1)$-dimensional metric,
and will determine which kind of BCFT is realized in AdS/BCFT.

In the Fefferman-Graham gauge 
the bulk metric becomes the following form:
\begin{equation}
ds^{2}=\frac{dz^{2}}{z^{2}}+g_{\mu\nu}dx^{\mu}dx^{\nu}.
\end{equation}
In this gauge the bulk metric admits a Taylor series expansion in
the neighborhood of the asymptotic boundary:
\begin{equation}
g_{\mu\nu}=g_{\mu\nu\left(0\right)}+z g_{\mu\nu\left(1\right)}+\cdots+z^{d-2}g_{\mu\nu\left((d-2)/2\right)}+\cdots,
\end{equation}
where the subscripts of the metric represent the order of the $z$ expansion. The boundary stress energy corresponds to the $g_{\mu\nu
\left((d-2)/2\right)}$
, so in terms of $g_{\mu\nu}$ we can rewrite as \cite{Banks:1998dd}:

\begin{equation}
T_{\mu\nu}=\underset{z\rightarrow0}{\lim}z^{-(d-2)}(g_{\mu\nu}-g_{\mu\nu}^{(0)} ), \label{eq:-4}
\end{equation}
where $g_{\mu\nu}^{(0)}$ is the AdS metric and we assumed 
there is no source term.

Then, for the tensionless ETW brane case, $T=0$, from the mode expansion \eqref{mode2}, 
we find $T_{\mu \nu}$ in the BCFT should satisfy \eqref{nbc} at the boundary $x=0$.\footnote{
In \cite{Suzuki:2022xwv} the authors mentioned the condition $T_{xi}=0$ in AdS/BCFT. }
This can be seen also from the linearized boundary condition (\ref{eq:21})-(\ref{eq:25}) directly.
Note that near $z=0$ any metric should be close to the AdS metric,
thus we can use the linearized analysis.
Thus, the BCFT${}_d$ realized in the AdS/BCFT with the tensionless ETW brane
can couple to the $d$-dimensional gravity with Neumann boundary condition.

For the nonzero tension ETW brane case, $T \neq 0$, we derive the constraints in subsection \ref{s3}, which are given by
\begin{equation}
    T_{ij}=-\frac{1}{d-1}\eta_{ij}T_{xx}.
\end{equation}
Thus, the BCFT${}_d$ realized in the AdS/BCFT with the nonzero tension ETW brane
can couple to the $d$-dimensional gravity with conformally Dirichlet boundary condition.
It will be a future problem to resolve why the Neumann like boundary condition in the AdS reduces to the conformally Dirichlet boundary condition in the BCFT.

 \section{Conclusion and Discussions}

In this paper we consider the AdS/BCFT correspondence for the pure gravity, especially in the vacuum AdS space. We analyze the perturbation of the metric and the boundary condition on the ETW brane in the sourceless case. 
In the tensionless case we find that the position of the ETW brane 
is fixed at the original hyperplane ($x=0$) in the FG gauge. 
We confirm that after imposing these boundary conditions the perturbation of the metric survives in the tensionless case. We also consider the boundary condition of the energy-momentum tensor in BCFT side and we find nontrivial constraints.
These conditions are necessary to define the BCFT which can  couple to a dynamical gravity with the Neumann or the conformally Dirichlet boundary conditions on the metric. 
We find that the BCFT${}_d$ realized in the AdS/BCFT with the tensionless ETW brane
can couple to the $d$-dimensional gravity with Neumann boundary condition.
For the AdS/BCFT with the nonzero tension ETW brane,
we find that 
the BCFT${}_d$
can couple to the $d$-dimensional gravity with conformally Dirichlet boundary condition.



\section*{Acknowledgement}
YS thank A.Ishibashi, Y.Okuyama, T.Ugajin and Z.Wei for helpful discussions. We also thank K.Izumi, T.Kawamoto, Y.Kusuki, T.Shiromizu, K.Suzuki, T.Takayanagi, N.Tanahashi, T.Tsuda and Y.Wang for giving useful comments on the draft of this paper and Y.Sato for  correcting our manuscript.
This work was supported by JSPS KAKENHI Grant Number 17K05414.
This work was supported by MEXT-JSPS Grant-in-Aid for Transformative Research Areas (A) ``Extreme Universe'', No. 21H05184.

\appendix
\section{Gauge invariant formalism}

\label{a1}

In this Appendix, we will use the gauge invariant formalism to study 
the perturbations of the AdS/BCFT instead of the FG gauge, however,
we find only a part of the mode expansion.

In preparation for the study of the dynamics of AdS/BCFT,
we will explicitly describe the 
free theory limit of gravity
in Poincare $AdS$ space, which is supposed to be dual to 
the generalized free limit of the holographic CFT on the Minkowski space.\footnote{
Assuming the BDHM relation \cite{Banks:1998dd},
the reconstruction of the bulk local operators can be done in principle.}
This will be done by knowing
the mode expansion of the gravitational
perturbation $g_{\mu\nu}^{\left(1\right)}$ 
around the $AdS_{n+3}$
background $g_{\mu\nu}^{\left(0\right)}$ in the Poincare coordinate:
\begin{equation}
ds^{2}=g_{\mu\nu}^{\left(0\right)} d x^\mu dx^\nu
=\frac{dz^{2}-dt^{2}+dx^{2}+\sum_{i=1}^{n} dw_{i}^{2}}{z^{2}},
\end{equation}
where $i=1,\ldots,n$. 
In this paper, we will assume $n \geq 1$ because for $n=0$ there is no usual gravitons
in $AdS_3$ and there are only the boundary gravitons. 
For the global $AdS$ case, this was already done in the gauge invariant
way 
in \cite{Ishibashi:2004wx}.
Thus, we will follow their study with a small number of modifications,
or just focusing the near boundary region to obtain the results for the Poincare patch.

For $x^\mu$, we will use latin indices in the range $a,b,\cdots$
to denote the components $\left(z,t\right)$ and also use the latin
indices in the range $p,q,\cdots,$ to denote the components $\left(x,w_{i}\right)$,
where $p,q, \cdots$ take $1, \dots, n+1$.
To denote the whole components we will use the greek indices. 
For the Poincare patch, 
a set of coordinates $\left(x,w_{i}\right)$, which describes ${\mathbf R}^{n+1}$,
has a rotational symmetry. 
Therefore the tensor $g_{ab}$ behaves
as the scalar field under this rotation. 
Similarly, $g_{ap}$ and $g_{pq}$
behave as the vector field and the symmetric tensor field, respectively. 
We can further decompose the vector field and the symmetric tensor field by 
the ``harmonics'' of ${\mathbf R}^{n+1}$  and their derivatives: 
\begin{eqnarray}
\mathbb{T_{\mathbf{k}}}_{pq}&=&\zeta_{pq}\exp(i k_p x^p),\nn\\
\mathbb{V_{\mathbf{k}}}_{p}&=&\zeta_{p}\exp(i k_p x^p),\nn\\
\mathbb{S_{\mathbf{k}}}&=&\exp(i k_p x^p),
\end{eqnarray}
where $k_p, \zeta_p, \zeta_{pq}=\zeta_{pq}$ are real constants such that $k^p \zeta_p=0, k^p \zeta_{pq}=0$ and $\sum_{p=1}^{n+1} \zeta_{pp}=0$.
These constants parameterize the harmonics.

Then, the metric
perturbations $g_{\mu\nu}^{\left(1\right)}$ can be decomposed into
$\left(z,t\right)$ directions and $\left(x,w_{i}\right)$ directions:

\begin{eqnarray*}
g_{ab}^{\left(1\right)}&=& h_{ab}^{\left(0 \right)}\\
g_{ap}^{\left(1\right)}&=&D_{p}h_{a}^{(0)} +h_{ap}^{\left(1\right)}\\
g_{pq}^{\left(1\right)}&=&h_{pq}^{\left(2\right)}+D_{p}h_{Tq}^{\left(1\right)}+D_{q}h_{Tp}^{\left(1\right)}\\
&+&h_{L}^{\left(0\right)}\gamma_{pq}+\left(D_{p}D_{q}-\frac{1}{n+1}\gamma_{pq} \gamma^{rs}D_{r}D_{s}\right)h_{T}^{\left(0\right)},
\end{eqnarray*}
where $D_p$ is the derivative in the Euclidean space and
\begin{eqnarray*}
\gamma_{pq}=\frac{g_{pq}}{z^{2}}.
\end{eqnarray*}
Here we note that the superscripts of the tensors $h$ denote the rank, not the order of the perturbation.  We also note that these tensors are defined such that
\begin{eqnarray*}
D^{p}h_{pq}^{\left(2\right)}=h_{p}^{\left(2\right)p}=0,\\
D^{p}h_{ap}^{\left(1\right)}=0,\\
D^{p}h_{Tp}^{\left(1\right)}=0.
\end{eqnarray*}
Thus, 
we can expand the tensor $h^{(a)}_{*} $ with $a=2,1,0$ by
$\mathbb{T_{\mathbf{k}}}_{pq},
\mathbb{V_{\mathbf{k}}}_{p},
\mathbb{S_{\mathbf{k}}}$, respectively and 
these are not mixed 
in the linearized Einstein equation.
We will call them tensor perturbation, vector perturbation and scalar perturbation,
respectively and the mode expansion can be done separately for these three perturbations.

\subsection*{Tensor perturbations}

We can expand the tensor type perturbation $h_{pq}^{\left(2\right)}$
in the basis $\mathbb{T_{\mathbf{k}}}_{pq}$ :
\begin{eqnarray}
h_{pq}^{\left(2\right)}=\underset{k_p, \zeta_{pq} 
}{\sum}H_{T}^{\left(2\right)}(z, t ; k_p, \zeta_{pq}) \, 
\zeta_{pq} e^{i k_p x^p},
\label{eq:40}
\end{eqnarray}
where the summations over $k, \zeta$ actually means the integrations. 
Since the gauge transformation 
\begin{eqnarray*}
g_{\mu\nu}\rightarrow g_{\mu\nu}+\nabla_{\mu}\xi_{\nu}+\nabla_{\nu}\xi_{\mu},
\end{eqnarray*}
does not contain tensor parts, 
$H_{T}^{\left(2\right)}$
is gauge-invariant. The equation of motion for this $H_{T}^{\left(2\right)}$ is the linearized Einstein equation,
\begin{equation}
\left(\delta_{\nu}^{\sigma}\nabla^{\rho}\nabla_{\mu}+\delta_{\mu}^{\sigma}\nabla^{\rho}\nabla_{\nu}-g^{\rho\sigma}\nabla_{\mu}\nabla_{\nu}-\delta_{\mu}^{\rho}\delta_{\nu}^{\sigma}\nabla^{2}-g_{\mu\nu}\nabla^{\rho}\nabla^{\sigma}+g_{\mu\nu}g^{\rho\sigma}\nabla^{2}\right)h_{\rho\sigma}=0,
\end{equation}
which implies that
the scalar 
$\Phi_{T}=z^{\frac{3-n}{2}}H_{T}^{(2)}$ satisfies 
\begin{equation}
\hat{\nabla_{c}}\hat{\nabla^{c}}(z^{\frac{3-n}{2}}\Phi_{T})-\left\{ \frac{(n+1)(n+3)}{4}+\left(\frac{n^{2}-1}{4} +\sum_{p=1}^{n+1} (k_p)^2
+2\right)z^{2}\right\} \Phi_{T}=0,
\end{equation}
where $\hat{\nabla_{c}}$ is the covariant derivative in the $AdS_2$ space 
parametrized by $z,t$.
This equation can be solved by the Bessel functions as 
\begin{equation}
H_{T}^{(2)}=\sum_\omega e^{i \omega t} \, z^{\frac{n-2}{2}}(a_\omega J_{\nu}(Z)+b_\omega Y_{\nu}(Z)),
\end{equation}
where we define 
\begin{equation}
\begin{array}{cc}
\nu=\frac{n+2}{2}, \, 
Z=\sqrt{ \omega^2 -\sum_{p=1}^{n+1} (k_p)^2 
} \, \, z.
\end{array}
\end{equation}

Because $J_{\nu}(Z) \sim Z^\nu$ and $Y_{\nu}(Z) \sim Z^{-\nu}$ near $z=0$,
the coefficients  $b_\omega=0$ so that the field should be (delta-function) normalized. 
Furthermore, we need to take $a_\omega=0$ if
$\omega^2 <\sum_{p=1}^{n+1} (k_p)^2$ because  $J_{\nu}(Z) \sim  e^{|Z|}$   in the limit
 $z \rightarrow \infty$.
Finally, the coefficients $a_\omega$ will be the creation operators or 
the annihilation operator for $\omega>0 $  or  $\omega < 0 $, respectively
with the suitable normalization constant.

\subsection*{Vector perturbations}

The vector components can be expanded in terms of the basis $\mathbb{V_{\mathbf{k}}}_{i}$:

\begin{eqnarray*}
h_{ai}^{\left(1\right)}=\sum H_{a}^{\left(1\right)}\zeta_{i}e^{ikx},\\
h_{Ti}^{\left(1\right)}=\sum H_{T}^{\left(1\right)}\zeta_{i}e^{ikx}.
\end{eqnarray*}
The gauge transformation is given by 

\begin{eqnarray*}
H_{a}^{\left(1\right)}\rightarrow H_{a}^{\left(1\right)}-\frac{1}{z^{2}}\nabla_{a}\left(z^{2}\xi^{\left(1\right)}\right),\\
H_{T}^{\left(1\right)}\rightarrow H_{T}^{\left(1\right)}-\xi^{\left(1\right)}.
\end{eqnarray*}
Then, we can define the gauge-invariant combinations

\begin{eqnarray*}
Z_{a}=H_{a}^{\left(1\right)}-\frac{1}{z^{2}}\hat{\nabla_a}\left(z^{2}H_{T}^{\left(1\right)}\right).
\end{eqnarray*}

To obtain the Bessel equation we define the scalar potential as
\begin{eqnarray*}
Z_a=z^{n-1}\epsilon_{ab}\hat{\nabla}^b\phi_V.
\end{eqnarray*}

Then the scalar potential $\phi_{V}$ satisfies 
\begin{equation}
\frac{1}{z^{n+3}}\hat{\nabla}_{c}(z^{n+1}\hat{\nabla}^{c}\phi_{V})-(k_{x}^{2}+k_{w_{i}}^{2}-n)\phi_{V}=0,
\end{equation}
and
we obtain the solution,
\begin{equation}
\phi_{V}=z^{-\frac{n}{2}}(a J_{\nu}(Z)+b Y_{\nu}(Z)),
\end{equation}
where we define 
\begin{equation}
\begin{array}{cc}
\nu=\frac{n}{2},\\
Z^{2}=(k_{x}^{2}+k_{w_{i}}^{2}-k^{2}+n)z^{2}.
\end{array}
\end{equation}

\subsection*{Scalar perturbations}

The scalar parts can be expanded in the scalar harmonics $S_{\mathbf{k}}$:

\begin{eqnarray*}
h_{ab}=\sum H_{ab}^{\left(0\right)}e^{ikx},\\
h_{a}=\sum H_{a}^{\left(0\right)}e^{ikx},\\
h_{L}=\sum H_{L}^{\left(0\right)}e^{ikx},\\
h_{T}^{\left(0\right)}=\sum H_{T}^{\left(0\right)}e^{ikx}.
\end{eqnarray*}
The gauge transformation can be summarized as

\begin{eqnarray*}
H_{ab}^{\left(0\right)}\rightarrow H_{ab}^{\left(0\right)}-\hat{\nabla_a}\xi_{b}^{\left(0\right)}-\hat{\nabla_b}\xi_{a}^{\left(0\right)},\\
H_{a}^{\left(0\right)}\rightarrow H_{a}^{\left(0\right)}-\xi_{a}^{\left(0\right)}-\frac{1}{z^{2}}\hat{\nabla_a}\left(z^{2}\xi^{\left(0\right)}\right),\\
H_{L}^{\left(0\right)}\rightarrow H_{L}^{\left(0\right)}+\frac{2k^{2}}{n+1}\xi^{\left(0\right)}-\frac{2}{z}\left(\hat{\nabla^a}\frac{1}{z}\right)\xi_{a}^{\left(0\right)},\\
H_{T}^{\left(0\right)}\rightarrow H_{T}^{\left(0\right)}-2\xi^{\left(0\right)}.
\end{eqnarray*}

We can construct the following gauge invariant combinations,

\begin{eqnarray*}
Z&=&\frac{(n+1)}{z^{(n-3)}}(H_L^{(0)}+\frac{k_s^2}{n+1}H_T^{(0)}+\frac{2}{z}(\hat{\nabla^a}\frac{1}{z})X_a)),\nn\\
Z_{ab}&=&\frac{1}{z^{(n-1)}}(H_{ab}^{(0)}+\hat{\nabla_a}X_b+\hat{\nabla_b}X_a)+\frac{n}{n+1}Zg_{ab},
\end{eqnarray*}
where
\begin{eqnarray*}
X_a=-H_a^{(0)}+\frac{1}{2z^2}\hat{\nabla_a}(z^2H_T^{(0)})
\end{eqnarray*}
The scalar potential $\Phi_{S}=z^{\frac{n+1}{2}}(\phi_{S}+\phi_{0})$
satisfies

\begin{equation}
\hat{\nabla_{c}}\hat{\nabla^{c}}\Phi_{S}-\left\{ \frac{(n-1)(n-3)}{4}+\left(\frac{n^{2}-1}{4}+k_{x}^{2}+k_{w_{i}}^{2}\right)z^{2}\right\} \Phi_{S}=0.
\end{equation}
We can solve this equation and the result is 

\begin{equation}
\phi_{S}+\phi_{0}=z^{-\frac{n}{2}}(a J_{\nu}(Z)+b Y_{\nu}(Z)),
\end{equation}
where we define 
\begin{equation}
\begin{array}{cc}
\nu=\frac{(n-2)}{2},\\
Z^{2}=(k_{x}^{2}+k_{w_{i}}^{2}-k^{2}+\frac{n^{2}-1}{4})z^{2}.
\end{array}
\end{equation}


We have obtained the free gravity theory around the $AdS_{n+3}$ space.
Here, we will study the CFT dual of this theory.
Let us consider the energy-momentum tensor $T_{\mu \nu}$,
where $\mu, \nu$ take $t, x^p$ only, in $CFT_{n+2}$. 
We expand $T_{pq} =\int_{\mathbf{k}, \omega} e^{i \omega t} (  a^T_{\mathbf{k} \omega} \, \mathbb{T_{\mathbf{k}}}_{pq}
+a^V_{\mathbf{k} \omega} \, D_q \mathbb{V_{\mathbf{k}}}_{p}
+a^S_{\mathbf{k} \omega} \, D_p D_q \mathbb{S_{\mathbf{k}}}
+a^{\rm Tr}_{\mathbf{k} \omega} \, \gamma_{pq} \,\mathbb{S_{\mathbf{k}}} )
$
where $\mathbf{k}$ include the polarization parameters $\zeta_p, \zeta_{pq}$.
The $T_{tt}, T_{t p}$ and $a^{\rm Tr}_{\mathbf{k} \omega} $ are not independent\footnote{
For $\omega=0$, $T_{tt}, T_{t p}$ are independent, however, 
$a^V, a^S$ are absent because $k_p=0$ 
with the condition $\omega^2 \geq \sum_{p=1}^{n+1} (k_p)^2$.}
{}from
$a^T, a^V, a^S$ because
the energy-momentum tensor should  satisfy the current conservation 
$0=\p^\mu T_{\mu \nu}=-\omega T_{t \nu} + k^p T_{p \nu}$ 
and traceless condition $0=T^\mu_{\,\mu} = - T_{tt} + \sum_{p=1}^{n+1} T_{pp}$.
Thus, 
if we assume the energy-momentum tensor behaves like the free theory
and only constraint by $\omega^2 \geq \sum_{p=1}^{n+1} (k_p)^2$.
which is the causality condition, except the current conservation and the traceless conditions,
it is equivalent to he free gravity theory around the Poincare $AdS_{n+3}$ space.
In principle, we can have the bulk reconstruction formula from this equivalence 
as for the scalar in the global $AdS$ space.

\subsection*{Gravitational perturbations in AdS/BCFT }

In this subsection, we consider the mode expansion of the 
gravitational perturbations in AdS/BCFT.
The mode expansion have been done for the case without the boundary condition  (\ref{eq:1}) in the previous section and then we will search which combinations of 
the modes satisfy the boundary condition, essentially.

In order to specify the boundary condition explicitly,  
we first write the background $AdS_{d+1}$ metric as,
 \begin{equation}
     ds^2=d\eta^2+\frac{\cosh{\eta^2}}{\zeta^2} (d\zeta^2-dt^2+\sum_{i=1}^{n} dw_{i}^{2})
 \end{equation}
where $n=d-2$,
which becomes
the familiar form of the metric in the Poincare coordinate,  
 \begin{equation}
ds^{2}=g_{\mu\nu}^{\left(0\right)} d x^\mu dx^\nu
=\frac{dz^{2}-dt^{2}+dx^{2}+\sum_{i=1}^{n} dw_{i}^{2}}{z^{2}},
\end{equation}
by the following coordinate transformation:
  \begin{eqnarray}
  z&=&\frac{\zeta}{\cosh{\eta}},\nn\\
  x&=&\zeta\tanh{\eta}.
  \end{eqnarray}
For this background, the 
boundary condition (\ref{eq:1}) is satisfied by restricting the spacetime
to the region $-\infty<\eta< \eta^*$,
where $\eta^*$ is determined by the tension $T$ as $T=(d-1) \tanh \eta^*$.
In the usual coordinate, the boundary is at 
$x/z=\sinh \eta^*$.
In particular, for tensionless case $T=0$, $\eta^*=0$ which means that the boundary is at  $x=0$.

Note that, If we choose the Gaussian normal coordinate like (\ref{eq:15}), 
the boundary condition (\ref{eq:1}) becomes just a partial derivative \cite{Nozaki:2012qd},
 \begin{equation}
     \frac{\partial h_{ab}}{dr}=\frac{2T}{d+1}h_{ab}=2\tanh{\eta_*}h_{ab}.\label{eq:17}
 \end{equation}

\subsection*{Tensionless case}

Let us consider the tensionless case.
The perturbation of the metric can be written as

\begin{eqnarray*}
g_{\mu\nu}=g_{\mu\nu}^{\left(0\right)}+g_{\mu\nu}^{\left(1\right)}.\\
\end{eqnarray*}
In this section we consider the Fourier transformed metric, but for simplicity we omit the label $k$ of the momentum. We will place a tensionless ETW at $x=0$ in this Poincare $AdS_{d+1}$
background $g_{\mu\nu}^{\left(0\right)}$. 
We further choose the coordinate system in which the ETW brane is at $x=0$
even after including the perturbations of the metric.

The normal vector to this
surface is

\begin{equation}
w=\left(\begin{array}{cc}
0\\
0\\
1\\
0\\
\vdots\\
0
\end{array}\right),
\end{equation}
where the coordinate were chosen to be $\{ t,z,x, w_i \} $.
Then, the normalized vector is given by 
\begin{eqnarray*}
n\simeq\left(\frac{1}{z}+\frac{z g_{xx}^{(1)}}{2}\right)\left(\begin{array}{c}
0\\
0\\
1\\
0\\
\vdots\\
0
\end{array}\right)
\end{eqnarray*}
up to the first order of the perturbation.
The extrinsic curvature is defined by
\begin{eqnarray*}
K_{ab}=h_{a}^{c}h_{b}^{d}\nabla_{c}n_{d},
\end{eqnarray*}
where the projection tensor is 
\begin{eqnarray*}
h_{a}^{c}=h_{ab}g^{bc}=\delta_{a}^{c}+\mathcal{O}\left(g^{\left(2\right)}\right).
\end{eqnarray*}
The Christoffel symbol in the linearized approximation is
\begin{eqnarray*}
\widetilde{\Gamma_{cd}^{e}}=\frac{1}{2}g^{\left(0\right)ef}\left(g_{fc,d}^{\left(1\right)}+g_{fd,c}^{\left(1\right)}-g_{cd,f}^{\left(1\right)}\right)+\frac{1}{2}g^{\left(1\right)ef}\left(g_{fc,d}^{\left(0\right)}+g_{fd,c}^{\left(0\right)}-g_{cd,f}^{\left(0\right)}\right).
\end{eqnarray*}
Substituting these altogether we get the linearized extrinsic curvature
\begin{eqnarray*}
K_{ab}=-\frac{1}{2}\delta_{a}^{c}\delta_{b}^{d}z\left(g_{xc,d}^{\left(1\right)}+g_{xd,c}^{\left(1\right)}-g_{cd,x}^{\left(1\right)}\right)-\frac{1}{2}\delta_{a}^{c}\delta_{b}^{d}\frac{g^{\left(1\right)xf}}{\sqrt{g^{\left(0\right)xx}}}\left(g_{xc,d}^{\left(0\right)}+g_{xd,c}^{\left(0\right)}-g_{cd,x}^{\left(0\right)}\right).
\end{eqnarray*}
Then, writing the boundary condition in components we find that
\begin{eqnarray*}
&K_{tt}=-z\left(g_{xt,t}^{(1)}-\frac{1}{2}g_{tt,x}^{(1)}\right)+\frac{g^{(1)xz}}{z^{4}}=0,\\
&K_{tz}=K_{zt}=-\frac{z}{2}\left(g_{xt,z}^{(1)}+g_{xz,t}^{(1)}-g_{zt,x}^{(1)}\right)-\frac{g^{(1)xt}}{z^{4}}=0,\\
&K_{zz}=-z\left(g_{xz,z}^{(1)}-\frac{1}{2}g_{zz,x}^{(1)}\right)+\frac{g^{(1)xz}}{z^{4}}=0,\\
&K_{tw}=K_{wt}=-\frac{z}{2}\left(g_{xw,t}^{(1)}+g_{xt,w}^{(1)}-g_{tw,x}^{(1)}\right)=0,\\
&K_{zw}=K_{wz}=-\frac{z}{2}\left(g_{xz,w}^{(1)}+g_{xw,z}^{(1)}-g_{zw,x}^{(1)}\right)-\frac{g^{(1)xw}}{z^{4}}=0,\\
&K_{w_{i}w_{i}}=-\frac{z}{2}\left(2g_{xw_{i},w_{i}}^{(1)}-g_{w_{i}w_{i},x}^{(1)}\right)-\frac{g^{(1)xz}}{z^{4}}=0,\\
&K_{w_{i}w_{j}}=g_{xw_{i},w_{j}}^{(1)}+g_{xw_{j},w_{i}}^{(1)}-g_{w_{i}w_{j},x}^{(1)}=0,
\label{eq:-6}
\end{eqnarray*}
where the extrinsic curvature was set to be zero since we focus on the tensionless
case. 

Next, we substitute the mode expansions of the metric perturbations obtained in the previous section which used \cite{Ishibashi:2004wx} to
the boundary conditions.
Below,  we will write the above conditions 
in the mode expansions
and omit 
the integrations over the momentum $k_p$ and the polarizations $\zeta_p , \zeta_{pq}$,
for notational simplicity.

The boundary condition $K_{tt}=0$ becomes
\begin{eqnarray*}
&e^{ikx}ik_{x}\left(\frac{\partial H_{t}^{(0)}}{\partial t}-\frac{1}{2}H_{tt}^{(0)}-\frac{H^{(0)}}{z}\right)=0,\\
&e^{ikx}\zeta_{x}\left(\frac{\partial H_{t}^{(1)}}{\partial t}-\frac{H_{z}^{(1)}}{z}\right)=0.
\end{eqnarray*}
Here  $e^{ikx}$ is the abbreviation of $e^{i k_p x^p} $.
Similarly for other components in the scalar part we get
\begin{eqnarray*}
&e^{ikx}ik_{x}\left(\frac{\partial H_{t}^{(0)}}{\partial z}+\frac{\partial H_{z}^{(0)}}{\partial t}-H_{zt}^{(0)}+\frac{2}{z}H_{t}^{(0)}\right)=0,\\
&e^{ikx}ik_{x}\left(\frac{\partial H_{z}^{(0)}}{\partial z}-\frac{1}{2}H_{zz}^{(0)}-\frac{1}{z}H_{z}^{(0)}\right)=0,\\
&e^{ikx}k_{x}k_{w}\frac{\partial H_{T}^{(0)}}{\partial t}=0,\\
&e^{ikx}k_{x}k_{w}\left(\frac{2}{z}H_{T}^{(0)}-\frac{\partial H_{T}^{(0)}}{\partial z}-2H_{z}^{(0)}\right)=0,\\
&e^{ikx}\left(-k_{x}k_{w}^{2}H_{T}^{(0)}-\frac{i}{2}k_{x}H_{L}^{(0)}-\frac{1}{2(n+1)}k_{x}^{2}H_{T}^{(0)}+\frac{n}{2(n+1)}k_{w}^{2}H_{T}^{(0)}+\frac{i}{z}H_{z}^{(0)}k_{x}\right)=0,\\
&e^{ikx}H_{T}^{(0)}k_{w_{i}}k_{w_{j}}k_{x}=0.
\end{eqnarray*}
For those in the vector part we find
\begin{eqnarray*}
&e^{ikx}\zeta_{x}\left(\frac{\partial H_{t}^{(1)}}{\partial t}-\frac{H_{z}^{(1)}}{z}\right)=0,\\
&e^{ikx}\zeta_{x}\left(\frac{\partial H_{t}^{(1)}}{\partial z}+\frac{\partial H_{z}^{(1)}}{\partial t}+\frac{2}{z}H_{t}^{(1)}\right)=0,\\
&e^{ikx}\zeta_{x}\left(\frac{\partial H_{z}^{(1)}}{\partial z}-\frac{H_{z}^{(1)}}{z}\right)=0,\\
&e^{ikx}\frac{\partial H_{T}^{(1)}}{\partial t}\left(ik_{x}\zeta_{w}+ik_{w}\zeta_{x}\right)-e^{ikx}H_{t}^{(1)}\left(ik_{x}\zeta_{w}-ik_{w}\zeta_{x}\right)=0,\\
&e^{ikx}\left[\left(ik_{x}\zeta_{w}+ik_{w}\zeta_{x}\right)\left(\frac{\partial H_{T}^{(1)}}{\partial z}-H_{T}^{(1)}\right)-\left(k_{x}\zeta_{w}-ik_{w}\zeta_{x}\right)H_{z}^{(1)}\right]=0,\\
&e^{ikx}\left[H_{T}^{(1)}\left(-k_{w_{i}}k_{x}\zeta_{w_{i}}-k_{w_{i}}^{2}\zeta_{x}+k_{x}k_{w_{i}}\zeta_{w_{i}}\right)+\frac{H_{z}^{(1)}}{z}\zeta_{x}\right]=0,\\
&e^{ikx}H_{T}^{(1)}e^{ikx}k_{w_{j}}k_{w_{i}}\zeta_{x}=0.
\end{eqnarray*}
Finally for the tensor part we obtain

\begin{eqnarray*}
e^{ikx}\zeta_{xw}\frac{\partial H_{T}^{(2)}}{\partial t}=0,\\
e^{ikx}\zeta_{xw}\left(\frac{\partial H_{T}^{(2)}}{\partial z}-\frac{2}{z}H_{T}^{(2)}\right)=0,\\
k_{w_{i}}\zeta_{xw}H_{T}^{(2)}=0,\\
e^{ikx}H_{T}^{(2)}\left(k_{w_{i}}\zeta_{xw_{j}}+k_{w_{j}}\zeta_{xw_{i}}-k_{x}\zeta_{w_{i}w_{j}}\right)=0.
\end{eqnarray*}

We can simplify the above constraints by the gauge transformation. 
For the scalar part we can set
$H_{a}^{\left(0\right)}=H_{T}^{\left(0\right)}=0$ since the gauge
transformation is linear. 
Secondly for the vector part we can set
$H_{T}^{\left(1\right)}=0$ and finally for the tensor part there
are no gauge transformations. 
Note that we already set the location of the ETW brane at $x=0$, which is 
not satisfied in a generic coordinate system with the perturbations.
Here, we will search the modes which satisfies the boundary conditions 
with the assumption  that the ETW brane is at $x=0$.
Then the above constraints can be simplified
as

\begin{eqnarray*}
&e^{ikx}ik_{x}\left(-\frac{1}{2}H_{tt}^{(0)}\right)=0,\\
&e^{ikx}ik_{x}\left(-H_{zt}^{(0)}\right)=0,\\
&e^{ikx}ik_{x}\left(-\frac{1}{2}H_{zz}^{(0)}\right)=0,\\
&e^{ikx}\left(-\frac{i}{2}k_{x}H_{L}^{(0)}\right)=0.
\end{eqnarray*}

\begin{eqnarray*}
&e^{ikx}\zeta_{x}\left(\frac{\partial H_{t}^{(1)}}{\partial t}-\frac{H_{z}^{(1)}}{z}\right)=0,\\
&e^{ikx}\zeta_{x}\left(\frac{\partial H_{t}^{(1)}}{\partial z}+\frac{\partial H_{z}^{(1)}}{\partial t}+\frac{2}{z}H_{t}^{(1)}\right)=0,\\
&e^{ikx}\zeta_{x}\left(\frac{\partial H_{z}^{(1)}}{\partial z}-\frac{H_{z}^{(1)}}{z}\right)=0,\\
&e^{ikx}H_{t}^{(1)}\left(ik_{x}\zeta_{w}-ik_{w}\zeta_{x}\right)=0,\\
&e^{ikx}\left[\left(k_{x}\zeta_{w}+ik_{w}\zeta_{x}\right)H_{z}^{(1)}\right]=0,\\
&e^{ikx}\left[\frac{H_{z}^{(1)}}{z}\zeta_{x}\right]=0,
\end{eqnarray*}

\begin{eqnarray*}
&e^{ikx}\zeta_{xw_{i}}\frac{\partial H_{T}^{(2)}}{\partial t}=0,\\
&e^{ikx}\zeta_{xw_{i}}\left(\frac{\partial H_{T}^{(2)}}{\partial z}-\frac{2}{z}H_{T}^{(2)}\right)=0,\\
&k_{w_{i}}\zeta_{xw_{i}}H_{T}^{(2)}=0,\\
&e^{ikx}H_{T}^{(2)}\left(k_{w_{i}}\zeta_{xw_{j}}+k_{w_{j}}\zeta_{xw_{i}}-k_{x}\zeta_{w_{i}w_{j}}\right)=0.
\end{eqnarray*}

The scalar part always
contains the factor $k_{x}$, which means $-i \p_x$.
Thus, by denoting the normalized mode without the boundary labeled by $\omega,k_x,k_{w_i}$ as
$g_{\mu\nu}^{(1) {\rm scalar}}(\omega,k_x,k_{w_i})$, the scalar mode with boundary at $x=0$ is given by
\begin{equation}
    \frac12 \left( g_{\mu\nu}^{(1) {\rm scalar}}(\omega,k_x,k_{w_i}) +g_{\mu\nu}^{(1) {\rm scalar}}(\omega,-k_x,k_{w_i}) \right),
\label{m1}
\end{equation}
which is proportional to $\cos (k_x x) $, i.e. satisfying the Neumann boundary condition.
Note that $H^{(0)}, H^{(1)},H^{(2)},$ are non-trivial functions of $z$ and $t$ even at $x=0$,
thus it is clearly impossible to get the mode satisfying the boundary condition
by imposing some condition to them, like $H^{(0)}_{tt}=0$.

For the vector part,  they are proportional to  $\zeta_{x}$ or $k_x$. 
Thus,  
by denoting the normalized mode without the boundary as
$g_{\mu\nu}^{(1) {\rm vector}}(\omega,k_x,k_{w_i},\zeta_x,\zeta_{w_i})$, the scalar mode with boundary at $x=0$ is given by
\begin{equation}
    \frac12 \left( g_{\mu\nu}^{(1) {\rm vector}}(\omega,k_x,k_{w_i},\zeta_x=0,\zeta_{w_i}) +g_{\mu\nu}^{(1) {\rm vector}}(\omega,-k_x,k_{w_i},\zeta_x=0,\zeta_{w_i}) \right).
\label{m2}
\end{equation}

For the tensor part 
they are proportional to  $\zeta_{x w_i}$ or $k_x$. 
Thus,  
by denoting the normalized mode without the boundary as
$g_{\mu\nu}^{(1) {\rm tensor}}(\omega,k_x,k_{w_i},\zeta_{x w_i},\zeta_{x x}, \zeta_{w_i w_j} )$, the scalar mode with boundary at $x=0$ is given by
\begin{equation}
    \frac12 \left( g_{\mu\nu}^{(1) {\rm tensor}}(\omega,k_x,k_{w_i},\zeta_{x w_i}=0,\zeta_{x x}, \zeta_{w_i w_j} ) +g_{\mu\nu}^{(1) {\rm tensor}}(\omega,-k_x,k_{w_i},\zeta_{x w_i}=0,\zeta_{x x}, \zeta_{w_i w_j} ) \right).
\label{m3}
\end{equation}

Let us summarize our results on the mode expansions of the AdS/BCFT for tensionless BTW brane.
They are given by \eqref{m1}, \eqref{m2} and \eqref{m3}, which are proportional to $\cos (k_x x) $ 
(the Neumann B.C. on the plane wave along $x$ direction)
and for which $\zeta_x=0$ and $\zeta_{x w_i}=0$.
Note that these are given in gauge invariant way because the mode expansions 
in \cite{Ishibashi:2004wx} is given in gauge invariant, although to obtain these we have chosen the particular gauge.
In the Gauss normal coordinate, the boundary condition seems to be given by the Neumann B.C. on the plane wave along $x$ direction only, however, we have chosen the different gauge and the mode expansions, which satisfy the linearized Einstein equation, are given in the gauge invariant way.\footnote{
These results are natural for $K_{ab}=0$ which is a ``Neumann B.C''.
For example, for the Maxwell theory on  four dimensional Minkowski space with the boundary at $x=0$, the  
``Neumann B.C'' is $F_{x \mu}=0$. Taking the $A_x=0$ gauge, 
the mode expansion of the gauge invariant field $E_i$ is given by $\zeta_i \cos{k_x} \, e^{i \sqrt{(k_x)^2+(k_y)^2+(k_z)^2} t+i k_y +i k_z z }$
where $\zeta_x=0$ and other gauge invariant field $B_i$ is expressed by $E_i$ essentially.

}

Note that for the vector and tensor part,
the degrees of freedom of the gravitational perturbations 
are less than the results in the FG gauge
because of the constraints $\zeta_x=0$ and $\zeta_{x w_i}=0$.
Thus, there are the modes which do not obey the assumption that 
the boundary is on $x=0$.
Because without the assumption, it seems difficult to find the explicit forms 
of the mode expansions, 
This problem is left for future work,

\subsection*{Nonzero tension case}
 
In this subsection we consider the nonzero tension case. In general it is hard to impose the boundary condition (\ref{eq:1}) explicitly. Therefore we choose the Gaussian normal coordinate like (\ref{eq:15}). Then the (\ref{eq:1}) becomes just a partial derivative \cite{Nozaki:2012qd},
 \begin{equation}
     \frac{\partial h_{ab}}{dr}=\frac{2T}{d+1}h_{ab}=2Th_{ab}.\label{eq:17}
 \end{equation}
 More precisely we can write the background AdS metric as,
 \begin{equation}
     ds^2=d\eta^2+\frac{\cosh{\eta^2}}{\zeta^2}(d\zeta^2-dt^2+dw_i^2),
 \end{equation}
where $-\infty<\eta<\eta^*$ and the coordinate transformation
  \begin{eqnarray}
  z&=&\frac{\zeta}{\cosh{\eta}},\nn\\
  x&=&\zeta\tanh{\eta},
\label{zx}
  \end{eqnarray}
  gives the familiar Poincare coordinate,
 \begin{equation}
     ds^2=\frac{dz^2+dx^2-dt^2+dw_i^2}{z^2}.
 \end{equation}
Then, the boundary condition \eqref{eq:17} for the perturbations around the background is 
\begin{equation}
\left.    \frac{\partial g^{(1)}_{ab}}{d \eta} \right|_{\eta=\eta^*}=
\left. 2 T g^{(1)}_{ab}  \right|_{\eta=\eta^*},.
\label{bc1}
 \end{equation}
where $a,b$ take $\{ \zeta, t, w_i \}$ and $T=\tanh \eta^*$.

In the $z,x$ coordinates, 
using \eqref{zx}, we find $\p_\eta=\frac{\zeta}{\cosh^2 \eta } \p_x 
- \frac{\zeta \sinh \eta}{\cosh^2 \eta } \p_z
 = \frac{1}{\cosh \eta } ( z \p_x -  x \p_z) $ and 
$x=z \sinh \eta^*$ at the boundary $\eta=\eta^*$.
Then, the  (\ref{bc1}). can be rewritten as 
 \begin{eqnarray}
& \left. \left(
 \frac{1}{\sinh^2 \eta^*} x \p_x -  z \p_z\right)  g_{ab}^{(1)}   \right|_{x=z \sinh \eta^*}=2 \left. g_{a b}^{(1)}\right|_{x=z \sinh \eta^*}.
\label{c1}
  \end{eqnarray}
Because the gauge invariant parts of the modes given before 
behave like $e^{k_x x} z^{d-2}$ near $z=0$,
the term including $\p_x$ in the \eqref{c1} are negligible near $z=0$ and $x=0$
if $ (z^{2-d} g_{a b}^{(1)} ) |_{x=z \sinh \eta^*} \neq 0 $ and 
we ignore the gauge dependent parts.\footnote{
In order to change this behavior, for example, to  $x^{d}$, we need to sum up infinitely large $k_x$, however, there is the constraint $\omega^2 \geq (k_x)^2$
which forbids such a linear combination with the fixed $\omega$. 
}
Then, \eqref{c1} becomes $ -  z \p_z  g_{w_iw_i}^{(1)}  |_{x=z \sinh \eta^*}=2 \left. g_{w_iw_i}^{(1)}\right|_{x=z \sinh \eta^*}$, 
which means $g_{w_iw_i}^{(1)}  |_{x=z \sinh \eta^*} \sim z^{-2}$.
This is for the non-normalizable mode, thus 
the gauge invariant parts of the modes alone can not satisfy the boundary condition.
The gauge dependent terms are determined by our  gauge choice here,
i.e. $g_{\eta M}^{(1)}=0 $.
Note that this gauge choice can be done before introducing the boundary
and we can impose $g_{\eta M}^{(1)}=0 $ by canceling  the contributions from the gauge invariant parts and the terms generated by the gauge transformation,
using the power series expansions by $ x$ and $z$.
This means that the dependent terms keep $z^{d-2}$ like behavior and can not 
have the $z^{-2}$ behavior.

Thus, for $T \neq 0$ there are no normalizable modes, 
the Dirichlet like boundary condition,
$ (z^{2-d} g_{a b}^{(1)} ) |_{x=z \sinh \eta^*} =0 $,
is required at the end point of the boundary in this gauge.

\end{document}